\documentclass[twocolumn,prl,nobalancelastpage,aps,10pt]{revtex4-1}
\usepackage{graphicx,bm,times}
\usepackage{textcomp}
\usepackage{natbib} 
\usepackage{etoolbox}

\begin{document}

\title{Quantum Tunneling Insights into the Atomic Landscapes of Graphite, Gold, and Silicon}

\author{Dhananjay Saikumar}

\affiliation{University of St Andrews}
\renewcommand{\thesubsection}{(\alph{subsection})}

\begin{abstract}
Scanning Tunneling Microscopy (STM) is a powerful technique that utilizes quantum tunneling to visualize atomic surfaces with high precision. This study presents detailed topographic maps and evaluates the local density of states (LDOS) for three distinct materials: Highly Oriented Pyrolytic Graphite (HOPG), gold, and silicon. By meticulously measuring the tunneling current from a finely pointed tip positioned nanometers above the sample, we successfully image the surface topography of HOPG, revealing a lattice constant of \(0.28 \pm 0.01\) nm. Additionally, we determine the local work functions for gold and graphite to be \(0.7 \pm 0.1\) eV and \(0.5 \pm 0.1\) eV, respectively. Employing scanning tunneling spectroscopy, this work further investigates the LDOS for gold (a metal), graphite (a semi-metal), and silicon (a semiconductor), providing valuable insights into their electronic properties at the atomic level.
\end{abstract}
\maketitle

\section{1. INTRODUCTION}

The invention of the Scanning Tunneling Microscope (STM) marked a revolutionary advancement in the field of surface science. Developed in 1981 by Heinrich Rohrer and Gerd Binnig at IBM's Zurich Research Laboratory, STM emerged from a curiosity to understand the local tunneling properties of oxide layers on metal surfaces at the nanoscale \cite{1}. The unveiling of atomic-scale images of a Si(111)-7x7 surface by Binnig et al. in 1982 showcased STM's unprecedented resolution, leading to the Nobel Prize in Physics for its inventors in 1986 \cite{2}.

At its core, STM is an electron microscope that leverages the principle of quantum tunneling, a phenomenon where, contrary to classical predictions, electrons have a nonzero probability of traversing a potential barrier even when their energy is insufficient. This quantum mechanical principle is elegantly exploited in STM by positioning a sharp metallic tip close to a sample and applying a bias voltage \textit{V}. The resulting tunneling current \textit{I} is exquisitely sensitive to the tip-sample separation \textit{z}, following the relationship derived from the one-dimensional tunneling problem \cite{3}:
\begin{equation}
I = I_{0} e^{-A\sqrt{\bar{\phi}}z}
\end{equation}
Here, \textit{I} represents the tunneling current, \textit{z} the separation gap, and $\bar{\phi}$ the average work function of the tip and sample, with \textit{A} being a constant derived from the Schrödinger equation for one-dimensional tunneling:
\begin{equation}
A = \frac{\sqrt{2m_{e}}}{\hbar} \approx 10.25 nm^{-1} eV^{-1/2}
\end{equation}

\textbf{Constant Current Imaging:} STM's capability to image surface topographies with atomic resolution, up to 0.1 nm ($1\AA$), is facilitated by maintaining a constant tunneling current. This is achieved through dynamic adjustment of the tip's position via piezoelectric transducers, regulated by a proportional integral derivative (PID) controller. Such precision enables STM to reveal intricate details of surface structures, positioning it as an indispensable tool for exploring the atomic and electronic landscapes of materials \cite{4}.

\section{2. Background}

\subsection{2.1 Imaging Techniques}

This study utilizes Scanning Tunneling Microscopy (STM) primarily for constant current imaging to explore the surface topography of Highly Oriented Pyrolytic Graphite (HOPG). Graphene, a carbon allotrope known for its strength and electrical conductivity, forms a hexagonal lattice that constitutes the layered structure of graphite. These layers are bound by weak Van der Waals forces and are arranged in an alternating A-B-A'-B' pattern, with the prime notation indicating distinct layers. This structural configuration results in differing atomic arrangements between A-A' and B-B' pairs, with the former having directly adjacent neighbors above and below. The lattice constant between A-A' pairs is measured at \(a = 0.246\) nm, while the atomic distance between A and B sites is \(c = 0.142\) nm, leading to the relationship \(a = \sqrt{3}c\). The regularity and periodicity of HOPG's structure make it an ideal candidate for STM imaging.

\setlength\belowcaptionskip{-1ex}
\begin{figure}[h]
  \includegraphics[width=.49\columnwidth]{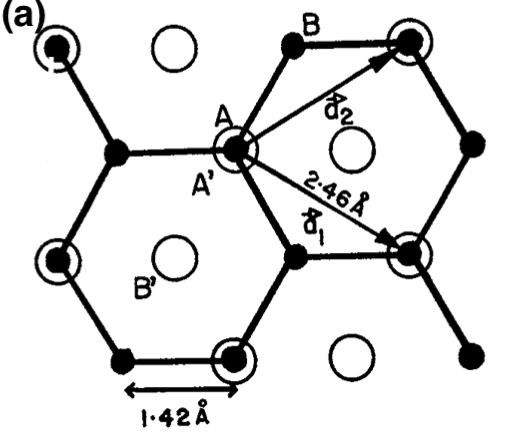} 
  \hfill  
  \includegraphics[width=.49\columnwidth]{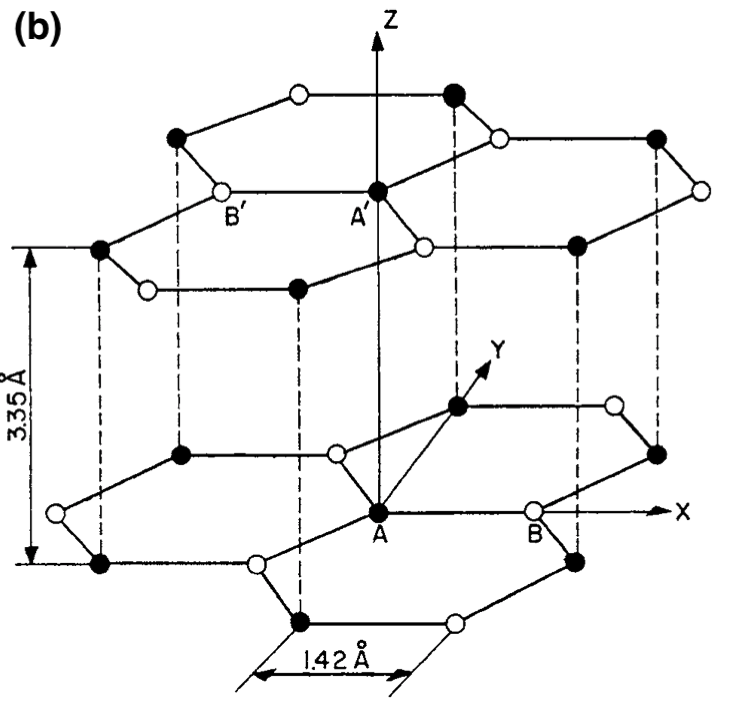}
\caption{Two-dimensional structure of graphite indicating ABA'B'
pattern on the left. Three-dimensional structure of graphite on the right is represented along with the lattice constant \textit{a} = 0.246 nm and atomic spacing \textit{c} = 0.142 nm, and atomic depth \textit{d} = 0.355 nm. Figure taken from  Ref. \cite{5}.}
\end{figure}

For the experimental setup, a meticulously sharpened gold tip, prepared using wire cutters, is integrated with the STM device. The HOPG sample is prepared for imaging by cleaving with scotch tape to expose a fresh surface. Utilizing the EasyScan software, provided by Nanosurf—a leading Swiss company in the development of AFM and STM technologies—the sample is brought into close proximity to the tip. The software controls the tip's movement across the x-y plane of the sample surface, adjusting its position dynamically to maintain a constant tunneling current, thereby enabling the detailed imaging of HOPG's atomic structure.

\subsection{2.2 The Work-Function}

The work-function, denoted as \(\phi\), plays a pivotal role in the study of electronic properties of materials. It represents the minimum energy required to extract an electron from the surface of a conductor to a point in vacuum immediately outside it \cite{7}. In the context of Scanning Tunneling Microscopy (STM), understanding the work-function is crucial as it directly influences the tunneling process. When a potential difference \(V\) is applied between the STM tip and the sample, electron tunneling occurs predominantly from states near the Fermi level \(E_f\), as these are the energy states available for electrons to transition into the vacuum.

The principle that the height of the potential barrier encountered by an electron in tunneling corresponds to the work-function (\(\phi = E_f - V\)) is central to interpreting STM data \cite{8}. The tunneling current \(I\), as described by the fundamental equation of STM, varies exponentially with the square root of the work-function and the separation gap \(z\), as shown below:
\begin{equation}
 \ln{I} = \ln{I_{0}} - A\sqrt{\bar{\phi}}z
\end{equation}
Here, the decay coefficient \(A\) and the average work-function \(\bar{\phi}\) between the tip and the sample are extracted from the slope of the linear relationship presented in the logarithmic form of the tunneling current equation.

Further analysis enables the calculation of the average work-function \(\bar{\phi}\) as:
\begin{equation}
\bar{\phi} = \frac{1}{A^2} \left( \frac{d\ln{I}}{dz} \right)^2
\end{equation}
This relationship provides a method to measure the work-function by employing STM in its spectroscopy mode, where the PID feedback loop is deactivated. In this mode, the tunneling current \(I\) is recorded as the tip is incrementally approached towards the sample surface (\(\Delta z\)) at a constant rate. By analyzing the exponential relationship between \(I\) and \(z\), along with the linear dependence of the logarithm of \(I\) on \(z\), the work-function \(\phi\) is determined. This approach offers insightful data on the electronic properties of the material under study, leveraging the nuanced capabilities of STM for surface analysis.

\subsection{2.3 The Local Density of States}

The tunneling current, \(I\), is modeled using time-dependent perturbation theory, specifically Bardeen's approach \cite{10}. This current is proportional to the integral of the product of the local density of states (LDOS) of the sample, \(\rho_{sample}\), and the LDOS of the tip, \(\rho_{tip}\), across the energy range up to the applied bias voltage \(eV\):
\begin{equation}
I \propto \int_{0}^{eV}\rho_{sample}(E_{f} - eV + \varepsilon) \rho_{tip}(E_{f} + \varepsilon) |M|^2 d\varepsilon
\end{equation}
Here, \(\rho_{sample}\) and \(\rho_{tip}\) denote the LDOS of the sample and tip, respectively, \(E_{f}\) is the Fermi energy, and \(|M|\) is the tunneling matrix element. For simplicity, \(\rho_{tip}\) and \(|M|\) are considered constants \cite{11}. This formulation leads to the conclusion that the differential conductance, \(\frac{dI}{dV}\), reflects the LDOS of the sample at the adjusted Fermi level:
\begin{equation}
\frac{dI}{dV} \propto \rho_{sample}(E_{f} - eV)
\end{equation}

To assess the LDOS, the STM operates in spectroscopy mode (Scanning Tunneling Spectroscopy, STS), where the tip's height above the sample is fixed. The variation in tunneling current \(I\) as a function of the bias voltage \(V\)—thereby shifting the Fermi level \(E_{f}\)—enables the calculation of the LDOS. The derivative of the current with respect to the voltage offers a profile of the LDOS.

The LDOS profile facilitates the classification of materials based on their electrical conductance. Metals, lacking a band gap, exhibit linear I/V curves indicating minimal DOS variation near \(E_{f}\). Semi-metals, with a narrow band gap, show non-linear I/V curves and a modest reduction in DOS near \(E_{f}\). In contrast, semiconductors, characterized by a significant band gap, display highly non-linear I/V curves, substantially diminishing the DOS around \(E_{f}\).

\begin{figure}[h]
\centering
\includegraphics*[width=0.6\linewidth,clip]{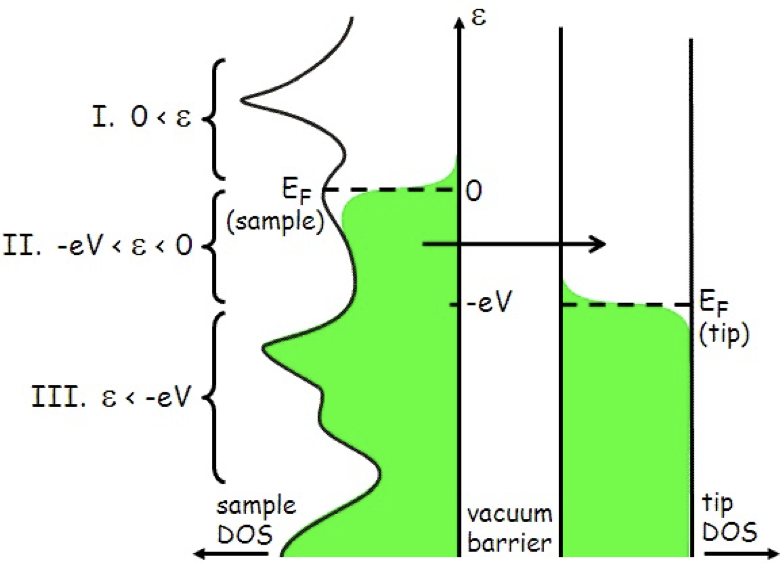}
\caption{Density of States (DOS) as a function of energy, with the applied bias voltage \(-V\) altering the Fermi energy \(E_{f}\) of the tip relative to the sample's \(E_{f}\). The shaded area indicates occupied states. This figure illustrates how scanning tunneling spectroscopy can derive a DOS profile by modulating \(-V\) and measuring the tunneling current \(I\). Source: \cite{12}.}
\end{figure}

\section{3. RESULTS \& DISCUSSION}

\subsection{3.1 Imaging}
The surface topography of Highly Oriented Pyrolytic Graphite (HOPG) was meticulously captured through constant current imaging with the Scanning Tunneling Microscope (STM). The imaging process involved acquiring multiple scans in the x-direction (fast-axis) and displacing them laterally in the y-direction (slow-axis), covering surface areas ranging from 1 to 4 \(nm^2\). These topographic scans unveiled a hexagonal lattice structure, characterized by two interlacing triangular sub-lattices, each comprising three carbon atoms.

\begin{figure}[h]
  \centering
  \includegraphics[width=.48\columnwidth]{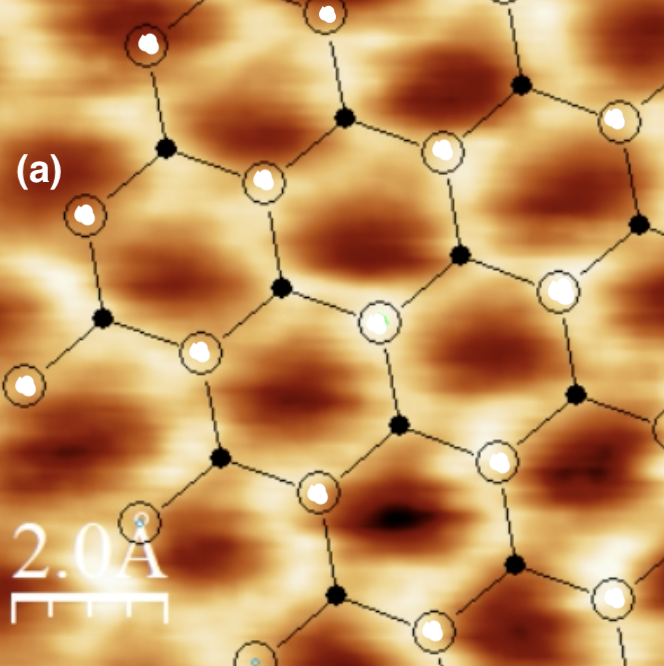} 
  \hfill  
  \includegraphics[width=.48\columnwidth]{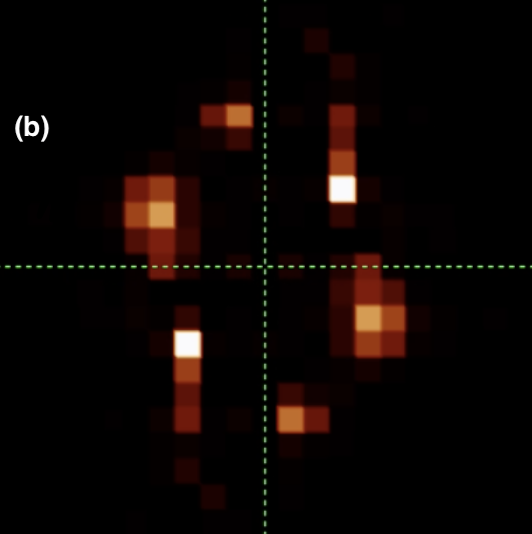}
\caption{(a) STM image of HOPG surface topography with a superimposed hexagonal lattice highlighting the graphite crystal structure. The hexagonal unit cell is marked by two basis atoms, A (white) and B (black), with dark rings denoting the hollow centers. (b) Fast Fourier Transform (FFT) of the image, representing the reciprocal space and elucidating the lattice's periodicity.}
\end{figure}

The hexagonal lattice overlay confirms the anticipated graphite structure, with the Fast Fourier Transform (FFT) of the 2D image dissecting the surface into its periodic components. Specifically, the FFT analysis showcases the electron plane wave's periodicity in the Bloch form, \(\psi_{k} = e^{ik\cdot r}u_{r}(c)\), with \(c\) representing atomic spacing.

By evaluating the spacing between alternating atoms in the reciprocal (K) space and employing the trigonometric relationship \(a = \sqrt{3}c\), the lattice constant was determined to average \(a = 0.28 \pm 0.01\) nm. This figure represents a 12.6\% increase over the commonly cited value of 0.246 nm in literature. Such deviation is attributed to image distortion along the slow-scan axis, a consequence of thermal drifts affecting the atoms and piezoelectric actuator hysteresis. The thermal drift, occurring at time scales shorter than the slow-axis scan, distorts the image similarly to how slow shutter speeds blur photographs.

The precision of the constant current feedback loop is also compromised by hysteresis, with asymmetries in the tip contributing further to image distortion. Electrochemical etching of the tip could mitigate these effects. Conducting experiments in a vacuum at reduced temperatures is recommended to minimize thermal drift and air molecule interactions. Historically, actuator-induced distortion was adjusted by averaging the drift effect with varying voltage. A novel post-correction method, Dynamic High-resolution Drift Tuning (DHDT), has been introduced, utilizing image features and nearest neighbors to develop a distortion model. This model compensates for thermal and hysteresis effects, with pixel corrections applied in inverse Fourier space \cite{15}.

\subsection{3.2 Work-function Measurement}

The local work-function, indicative of the barrier height, is derived from the logarithmic variation of the tunneling current \(I\) with the gap separation \(z\), as outlined in equation (3). The formula for calculating the average work function, \(\bar{\phi}\), is as follows:
\begin{equation}
\bar{\phi} = \frac{\phi_{Tip} + \phi_{Sample}}{2}
\end{equation}
Utilizing this approach, the average work functions of gold and graphite were measured with a gold tip. For the gold tip against a gold sample (\(Au_{tip}\)--\(Au_{sample}\)), it is established that \(\bar{\phi} = \phi_{Au}\). This value, combined with the measured \(\bar{\phi}\) from a gold tip and graphite sample (\(Au_{tip}\)--\(Graphite_{sample}\)), allows for the determination of the graphite work-function:
\begin{equation}
\phi_{Graphite} = 2\bar{\phi} - \phi_{Au}
\end{equation}
Measurements conducted across a 100 \(nm^2\) surface area revealed a linear relationship between the logarithmic I vs Z spectra, in line with theoretical expectations. The gradient of this linear relationship was ascertained through a straightforward linear fit.

\begin{figure}[h]
\centering
\includegraphics[height=0.65\linewidth,width=\linewidth,clip]{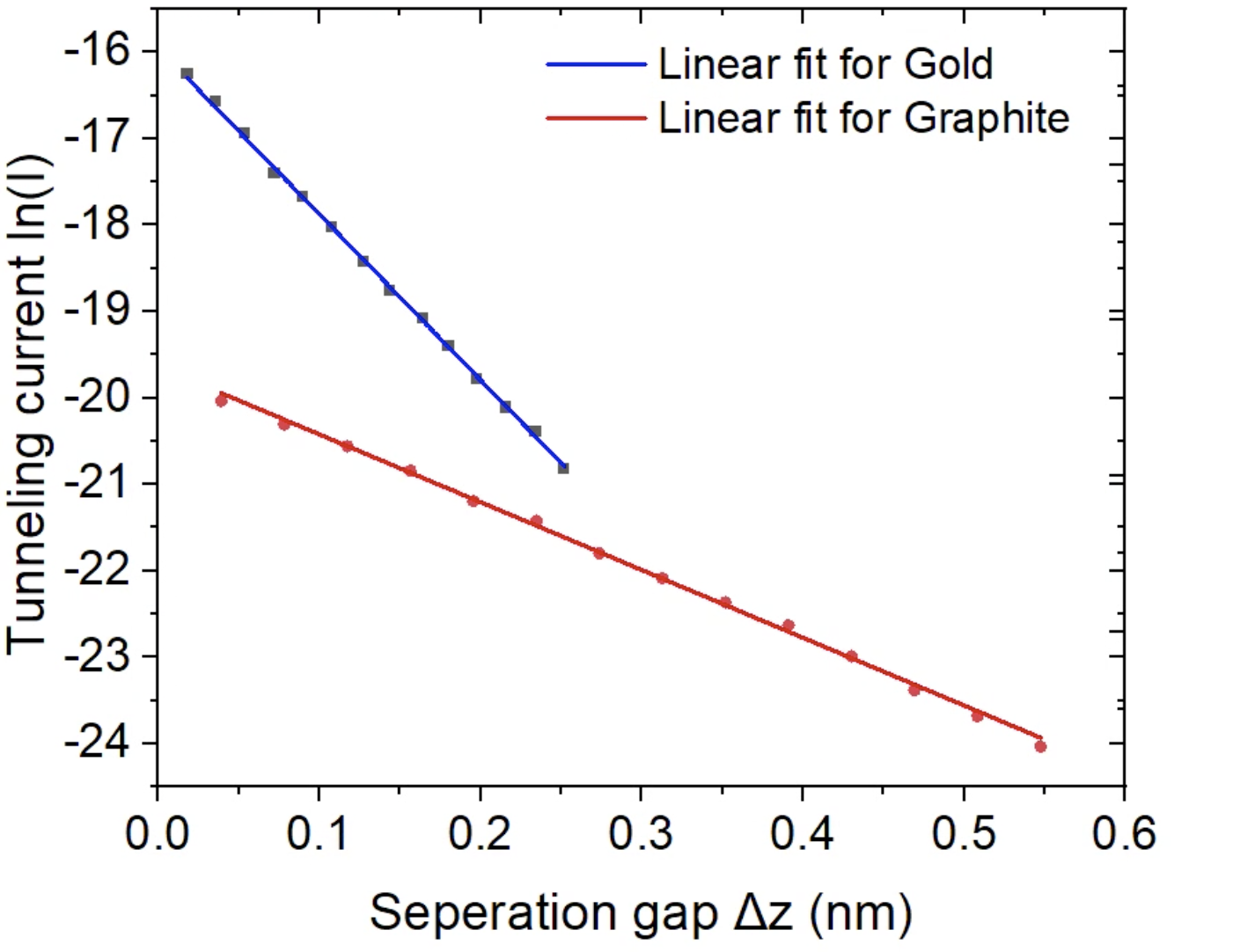}
\caption{Logarithmic variation of current \(I\) with tip separation \(z\) for Gold (blue) and Graphite (red) samples, maintained at a bias voltage of 50 mV and a temperature of 24.4\textdegree{}C. The extrapolated work-functions are \(\phi_{Gold} = 0.83\) eV, and \(\bar{\phi} = 0.63\) eV.}
\end{figure}

Averaging over ten measurements, the work-function for the gold sample was found to be \(\phi_{Gold} = 0.7 \pm 0.1\) eV, showing an 80\% deviation from the literature value of 5.32 eV \cite{16}. Similarly, ten measurements on the HOPG sample yielded an average work-function of \(\bar{\phi} = 0.7 \pm 0.1\) eV. By applying equation (8), the work-function of graphite is calculated to be \(\phi_{Graphite} = 0.5 \pm 0.1\) eV, diverging by 84\% from the literature value of 4.62 eV \cite{17}.

The slope comparison between gold and graphite lines in Figure 4 highlights a higher work-function for gold, consistent with literature values. The substantial discrepancies from literature work-functions suggest several influencing factors, including the angle \(\alpha\) between the tip and the sample, affecting the work-function by a factor of \(cos^2\alpha\) \cite{18}. Deviations observed in non-vacuum tunneling experiments, where measured work-functions did not exceed 1 eV, further corroborate the impact of environmental conditions on measurements \cite{19}.
The Smoluchowski effect describes how structural defects and the resulting non-uniform electron distribution create regions of lower potential, thus facilitating electron extraction \cite{20}. An alternative method for determining work-functions, observing Gundlach oscillations via STS, highlights work-functions through field-emission resonances~\cite{21}.

\subsection{3.3 Local Density of States Analysis}

\begin{figure}[h]
\centering
\includegraphics[height=0.65\linewidth,width=\linewidth,clip]{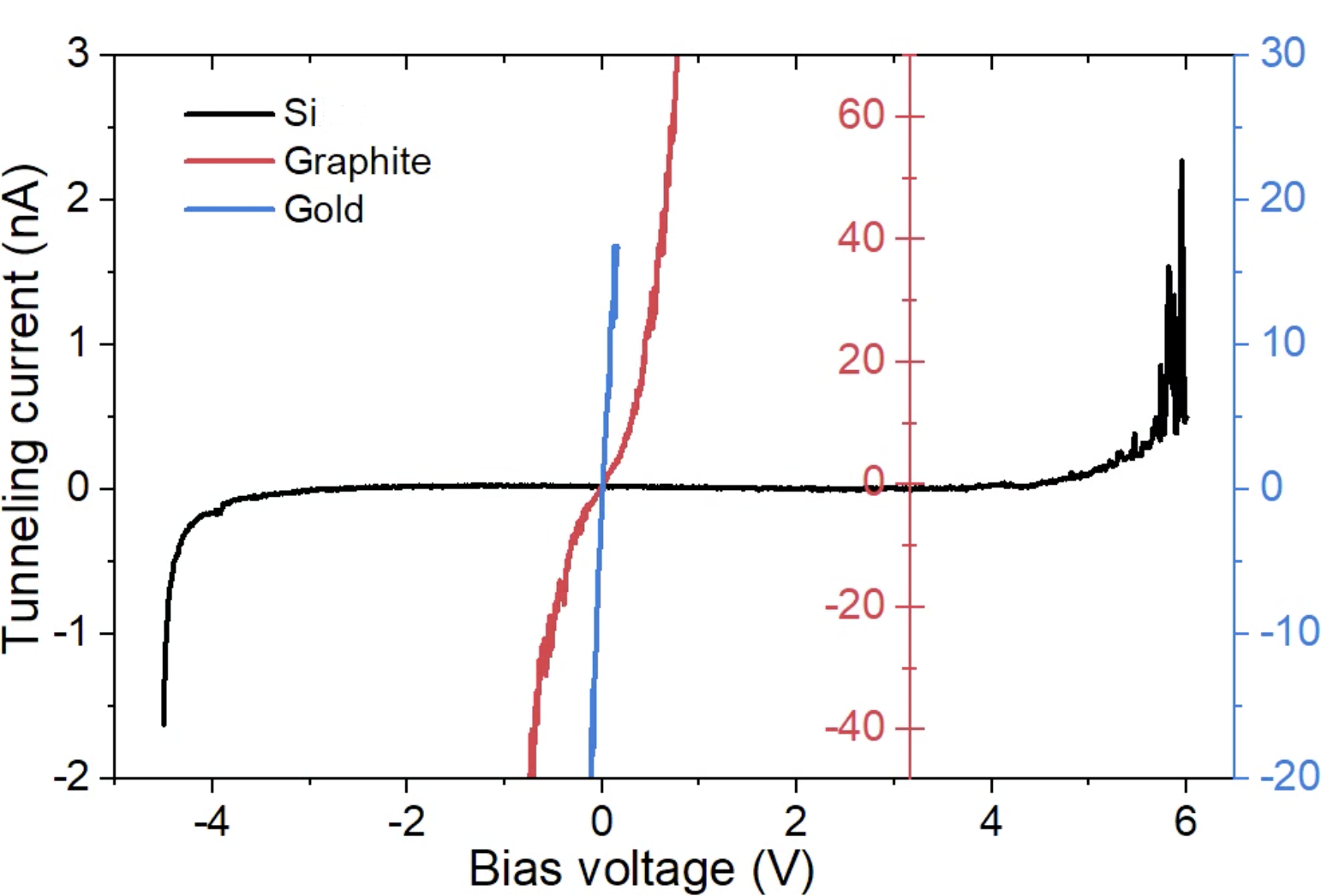}
\caption{I vs V spectra for Gold (blue), Graphite (red), and Silicon (black) measured at 21\textdegree{}C, highlighting the noise in the silicon spectra beyond a bias voltage of +4.3V.}
\end{figure}

The LDOS profiles for gold, graphite, and silicon were determined through numerical differentiation of their respective I/V spectra. This differentiation process introduces noise, particularly pronounced at higher bias voltages. Consequently, the LDOS profiles are analyzed within a confined voltage range of -0.04 mV to 0.04 mV to mitigate the effects of noise. Figure 6 illustrates these profiles, facilitating the comparison of the materials based on their conductance in descending order: gold, graphite, and silicon.

Typically, LDOS measurements extend over a broader interval (from -5 V to 5 V) to provide a comprehensive depiction of the LDOS, revealing band edges and enabling the determination of the band gap. However, as demonstrated in Figure 5, expanding the measurement range introduces significant noise. To counteract this, employing a lock-in amplifier to modulate the voltage sinusoidally—superimposed onto the bias voltage—significantly enhances measurement sensitivity and accuracy \cite{22}.

\begin{figure}[h]
\centering
\includegraphics[height=0.65\linewidth,width=\linewidth,clip]{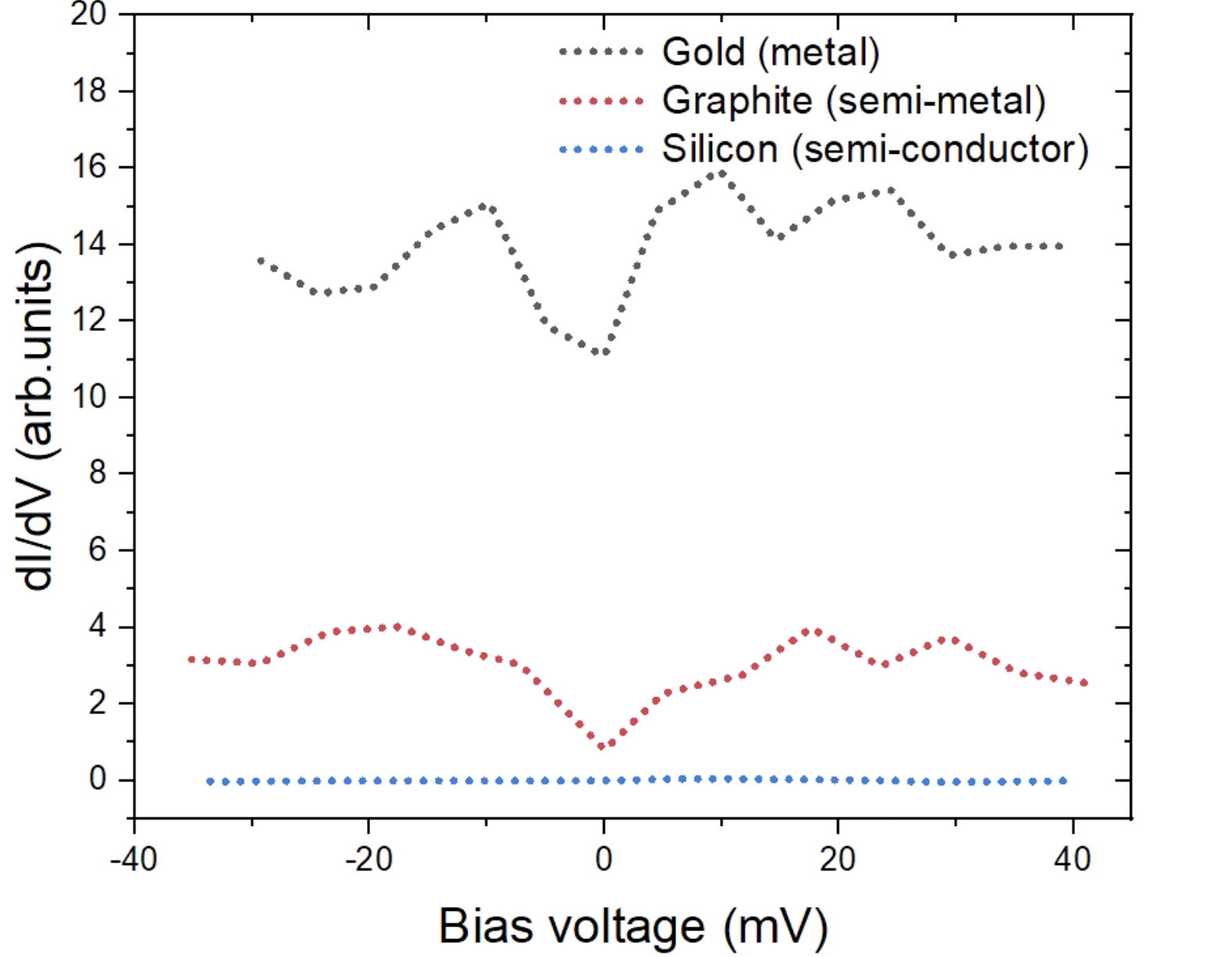}
\caption{Narrow-range LDOS profile for Gold (gray), Graphite (red), and Silicon (blue) from -0.04 mV to 0.04 mV. Gold and Graphite exhibit non-zero conductance, contrasting with Silicon's zero conductance due to a band gap of 1.6 eV.}
\end{figure}

\section{4. CONCLUSIONS}

Scanning Tunneling Microscopy (STM) has proven to be an invaluable tool for analyzing surface topography at the atomic level, as demonstrated by the detailed imaging of Highly Oriented Pyrolytic Graphite (HOPG). This study yielded a lattice constant of \(0.28 \pm 0.01\) nm for HOPG. Additionally, the local work-functions for gold and graphite were determined to be \(0.7\) eV and \(0.5\) eV, respectively, utilizing the I vs Z spectral analysis. Further, Scanning Tunneling Spectroscopy (STS) facilitated the acquisition of the Local Density of States (LDOS) maps for gold, graphite, and silicon, highlighting distinct electronic properties.

A significant recommendation for enhancing the accuracy of measured parameters, such as the lattice constant and work-function, is conducting experiments in a vacuum at very low temperatures. A vacuum environment minimizes thermal distortions by eliminating air interactions, potentially increasing the measured work-function by removing the insulating effect of air. Lowering the temperature mitigates thermal drift, further improving the precision of these measurements.

\section{references}


\begin{thebibliography}{99}

\bibitem{1} \textit{Scanning tunneling microscopy}, G. Binnig, H. Rohrer,  IBM Journal of Research and Development, (1982).

\bibitem{2} \textit{7 x 7 Reconstruction on Si(111) Resolved in Real Space}, G. Binnig, H. Rohrer, Physical Review Letters \textbf{50}, (1983).

\bibitem{3} \textit{Scanning tunneling microscopy}, Hansma, P. K., \& Tersoff, J, Journal of Applied Physics \textbf{62}, 340, (1988).

\bibitem{4} \textit{Origin of Atomic Resolution on Metal Surfaces in Scanning Tunneling Microscopy}, Chen, C. Julian, Phys. Rev.  \textbf{65}, (1990).

\bibitem{5} \textit{Review Graphite}, D. Chung, Journal of Material Sci \textbf{37}, (2002).

\bibitem{6} \textit{The structure of graphite}, Bernal John Desmond and Bragg William Lawrence, \textbf{106} Proc. R. Soc. Lond. A , (1924).

\bibitem{7} \textit{Introduction to Solid State Physics (7th ed.)}, Kittel. Charles, - Wiley (2007).
 
\bibitem{8} \textit{Theory of the scanning tunneling microscope}, Tersoff, J. and Hamann, D .R, Phys. Rev. B Vol \textbf{31}, (1985).
 
\bibitem{9} \textit{Introduction to scanning tunneling microscopy}, Julian Chen. Oxford University Press, page \textbf{5}, (2008).

\bibitem{10} \textit{Theory of scanning tunneling spectroscopy}, Julian Chen. IBM Thomas J. Watson Research Center, page \textbf{5}, (1987).
 
\bibitem{11} \textit{Local density of states from spectroscopic STM}, Li, J., Schneider, W.-D, Berndt, R, Phys. Rev. B Vol \textbf{56}, (1997).
 
 \bibitem{12} \textit{Scanning tunneling microscopy}, Hoffman Lab. (2010). [online]  Available at:
 hoffman.physics.harvard.edu/STMtechnical.php.
 
 \bibitem{13} \textit{Fourier Transform–STM: determining the surface Fermi contour}, L. Petersena, Ph. Hofmannb, E.W. Plummerc,d, F. Besenbachera, Journal of electron spectroscopy and related phenomena, 109 \textbf{1-2}, (1999).
 
 \bibitem{14} \textit{Tunneling microscopy study of the graphite surface in air and water}, J. Schneir, R. Sonnenfeld, P. K. Hansma, and J. Tersoff,Phys. Rev. B \textbf{34}, (1986).

 \bibitem{15} \textit{Post-processing correction of thermal drift and piezoelectric actuator nonlinearities in STM images}, Mitchell P. Yothers, Aaron E. Browder, and Lloyd A. Bumm. Review of Scientific Instruments \textbf{88}, 013708 (2017).

\bibitem{16} \textit{Work function of metals. Solid Surface Physics}, Hölzl, J., \& Schulte, F. K., \textbf{1-150}, (1979).

 \bibitem{17} \textit{The Thermionic Constants of Metals and Semi-Conductors.}, Jain, S. C., \& Krishnan, K. S., R, Royal Society \textbf{(213)}, (1952).

 \bibitem{18} \textit{Apparent tunneling barrier height and local work function of atomic arrays}, Neda Noei, Alexander Weismann, Richard Berndt, Beilstein J. Nanotechnol, (2018).
 
  \bibitem{19} \textit{Scanning tunneling microscopy - from birth to adolescence}, Binnig, G.; Rohrer, H. IBM J. Res. Dev. \textbf{(30)}, 355, (1986).
  
 \bibitem{20} \textit{Anisotropy of the Electronic Work Function of Metals}, R. Smoluchowski: Phys. Rev. \textbf{(60)}, 661, (1941). 
 
 \bibitem{21} \textit{Solid State Electron}, K. H. Gundlach, Vol \textbf{(9)}, (1996).

 \bibitem{22} \textit{Scanning tunneling spectroscopy of high-Temperature superconductors}, O.Fischer, C.Berthod,: Rev. Mod. Phys \textbf{(79)}, (2007).

\end{thebibliography}
\end{document}